\documentstyle[preprint,prb,aps,epsf]{revtex}

\newcommand{\be}{\begin{equation}}
\newcommand{\ee}{\end{equation}}
\newcommand{\bea}{\begin{eqnarray}}
\newcommand{\eea}{\end{eqnarray}}
\newcommand{\ba}{\begin{array}}
\newcommand{\ea}{\end{array}}
\newcommand{\vp}{\varphi}
\newcommand{\ep}{\epsilon}
\newcommand{\Ga}{\Gamma}
\newcommand{\ga}{\gamma}
\newcommand{\al}{\alpha}
\newcommand{\ka}{\kappa}
\newcommand{\la}{\lambda}
\newcommand{\de}{\delta}

\newcommand{\om}{\omega}

\newcommand{\br}{{\bf r}}
\newcommand{\Si}{\Sigma}
\newcommand{\ta}{\theta}
\newcommand{\tac}{\theta^{\ast}}
\newcommand{\zc}{z^{\ast}}
\newcommand{\im}{\mbox{Im}}
\newcommand{\re}{\mbox{Re}}
\newcommand{\cD}{{\cal D}}
\newcommand{\cA}{{\cal A}}

\newcommand{\ub}{\bar{u}}
\newcommand{\vb}{\bar{v}}
\newcommand{\Phib}{\bar{\Phi}}
\newcommand{\vpb}{\bar{\varphi}}

\newcommand{\chib}{\bar{\chi}}
\newcommand{\bep}{\bar{\epsilon}}

\newcommand{\xic}{\xi^{\ast}}
\newcommand{\phic}{\phi^{\ast}}
\newcommand{\etac}{\eta^{\ast}}

\begin{document}
\title{Localized states in strong magnetic field: resonant
scattering and the Dicke effect}
\author{T. V. Shahbazyan\footnote{Current address: Department of
Physics and Astronomy, Vanderbilt University, Nashville, TN 37235} 
and S. E. Ulloa}
\address{Department of Physics and Astronomy,
Condensed Matter and Surface Science Program, Ohio University,
Athens, OH 45701-2979}
\maketitle
\draft
\begin{abstract}
 We study the energy spectrum of a system of localized states coupled
to a 2D electron gas in strong magnetic field. If the energy levels of
localized states are close to the electron energy in the plane, the
system exhibits a kind of collective behavior analogous to the Dicke
effect in optics.  The latter manifests itself in ``trapping'' of
electronic states by localized states. At the same time, the
electronic density of states develops a gap near the resonance. The gap
and the trapping of states appear to be complementary and reflect an
intimate relation between the resonant scattering and the Dicke
effect. We  reveal this relation by presenting the {\em exact}
solution of the problem for the lowest Landau level. 
In particular, we show that in the absence of disorder the system
undergoes a phase transition at some critical concentration of
localized states.
\end{abstract}
\narrowtext

\section{introduction}

Electronic states of two--dimensional (2D) systems in a magnetic field
in the presence of impurities have been intensively studied during
the last two decades.\cite{A,and74,B,D,E,F,G,I,J,K,L,M,Q,P,hajdu,T}  The
macroscopic degeneracy of the Landau levels (LL) makes impossible a
perturbative treatment of even weak disorder and calls for
non--perturbative approaches. For high LL, Ando's self--consistent Born
approximation\cite{A} was shown to be asymptotically exact for
short--range disorder,\cite{I,P} while in the case of long--range
disorder the averaged density of states (DOS) was obtained within
the eikonal approximation.\cite{P} For low LL and uncorrelated
disorder, the problem contains no small parameter and neither of those
approximations applies. Nevertheless, for the lowest LL, the exact DOS in
a white--noise potential has been obtained by Wegner by mapping the
problem onto that of the 0D complex $\phi^4$--model.\cite{F} This
remarkable result  was extended to non--Gaussian distributions of
random potential by  Br\'ezin, Gross, and Itzykson within the
functional--integral    approach,\cite{G} and recently to multilayer
systems.\cite{T}

In the works mentioned above, the energy levels of the impurities
played no role in the scattering. Experimentally, this is well
justified since usually the random potential comes from the charged
donors with energy levels substantially higher than the Fermi energy in
the plane. The Gaussian form of the distribution function implies that
random potential is created by a large number of relatively 
weakly scattering impurities. The LL shape is then described by a
``smooth'' curve, symmetric with respect to the LL center.  In the case
of point--like scatterers with constant scattering strength, the DOS is
strongly asymmetric,\cite{and74,B,G,J}  vanishing below (above) the LL
center for repulsive (attractive) potential. An asymmetry, caused by
deviations from the Gaussian distribution, has been observed in very
low--mobility heterostructures.\cite{hau87}

The situation is quite different in the presence of localized states
(LS) with energies close to the electron energy in the plane.  A large
number of such LS's  can have a dramatic effect on
the properties of the 2D electron gas.  Such experimental structures
became available with recent advances in fabrication of arrays  of
ultra--small InAs  quantum dots with unusually narrow distribution of
parameters.\cite{leo93,mar94,gru95}  With typical sizes of less than
$20$ nm and variations of less that 10\%, an array of such dots with
density $10^{10}-10^{11}$ cm$^{-1}$ can be produced at some preset
distance from a plane of high mobility electrons.\cite{sak95} As the
Fermi energy in the plane is brought close to the levels of dots, the
scattering becomes strongly enhanced. It was, in fact, observed in
Ref.\onlinecite{sak95} that the mobility dropped by two orders of
magnitude when the thickness of the tunneling barrier between the dots
and the plane was reduced. 

In this paper we study the electronic states of a system consisting of
2D electron gas in strong magnetic field and point--like LS's with energy
levels close to the electron energy in the plane. It is important to
realize that the {\em combined} effect of such  LS's differs
drastically from that of a collection of isolated LS's. The reason lies
in a specific type of coupling between LS's, which  originates
from electronic transitions between LS's and the electron plane. For an
isolated LS, such transitions merely lead to  broadening of the LS
level. However, in the case of many LS's, the electron in the
``course'' of a single transition between a particular LS and the the
plane, ``visits'' also the {\em other} LS's, propagating in the plane
between successive transitions. As a result, the LS's become coupled via
the states in the plane. This coupling differs qualitatively from the
usual overlap of the LS wave--functions, and leads to formation of
certain {\em coherent} state.\cite{sha94}  

Let us illustrate the role of such coupling between LS's (in the
absence of magnetic field) on the following example. Consider first an
isolated LS with energy $\ep_1$. In the absence of tunneling, the spectral 
function (SF) of the LS is simply $S_0(\om)=\de(\om-\ep_1)$. Turning on
the tunneling transforms the SF into the Lorentzian 

\be\label{Siso}
S(\om)=-{1\over\pi}\im{1\over \om-\ep_1+iW}
={1\over\pi}{W\over (\om-\ep_1)^2+W^2},
\ee
with 

\be\label{Giso}
W=\pi\sum_{{\bf k}}|t_{1{\bf k}}|^2
\delta(\om-E_{{\bf k}}),
\ee
where $t_{1{\bf k}}$ is the amplitude of tunneling between the
LS and state ${\bf k}$ in the plane. The meaning of Eq.~(\ref{Siso})
is, of course, that in the presence of tunneling the LS level
acquires a width, $W^{-1}=\tau$ being the decay time (we set $\hbar=1$).  

Let us now place another LS with energy $\ep_2$ at some distance from
the LS 1. Then a simple generalization of Eq.~(\ref{Siso}) gives 

\be\label{Spair}
S(\om)=-{1\over\pi}\im{1\over 2}
{\rm Tr}{1\over \om-\hat{\ep}+i\hat{W}},
\ee
where $\hat{\ep}$ is diagonal $2\times 2$ matrix with eigenvalues
$\ep_1$ and $\ep_2$, and 

\be\label{Gpair}
W_{ij}=
\pi\sum_{{\bf k}}t_{i{\bf k}} t_{{\bf k} j}\delta(\om-E_{{\bf k}})
\ee
is the matrix of widths. The key observation is that the matrix elements of 
$\hat{W}$ are not independent, but instead satisfy a certain
relation.\cite{sha94} This relation follows from the definition of   
$t_{i{\bf k}}$ as the overlap between wave functions of the LS $i$ and
of the state ${\bf k}$ in the plane. Since the latter is simply a
plane wave, $t_{i{\bf k}}$ contains a phase factor depending on the
in-plane coordinate ${\bf r}_i$ of the LS:   
$t_{i{\bf k}}=e^{i{\bf k \cdot r}_i}t_i$, with
$|{\bf k}|=k_{_F}=2\pi/\la_{_F}$, $\la_{_F}$ being the Fermi
wave length. For diagonal elements, $W_{ii}\equiv W_i$, the product
$t_{i{\bf k}} t_{{\bf k} i}$ in Eq.~(\ref{Gpair}) is 
independent of orientation of ${\bf k}$. For non--diagonal elements,
however, the product $t_{i{\bf k}}t_{{\bf k} j}$  
contains the factor $e^{i{\bf k} \cdot {\bf r}_{ij}}$, $r_{ij}$ being the 
distance between the LS's. One then obtains from 
Eq.~(\ref{Gpair})

\be\label{W}
W_{12}=q\sqrt{W_{1}W_{2}},~~~
q=J_0(r_{12}k_{_F}),
\ee
where $J_0(x)$ is the Bessel function. For the simplest case of identical
LS's, $\ep_i=\ep$, $W_i=W$ and $W_{12}=qW$, Eq.~(\ref{Spair}) yields 

\be\label{Spair1}
S(\omega)={1\over \pi}
\Biggl[{1\over 2}{W_{-}\over (\omega-\ep)^2+W_{-}^2}
+{1\over 2}{W_{+}\over (\omega-\ep)^2+W_{+}^2}\Biggr],
\ee
with $W_{\pm}=(1\pm q)W$. 

It can be seen that if the two LS's are well separated,
$r_{12}k_{_F}\gg 1$, then the parameter $q$ is small and the SF 
again has the form of simple Lorentzian with 
the width $W$. However, if the distance between LS's is smaller than
the Fermi wave length, $r_{12}k_{_F}\lesssim 1$, then
$q\sim 1$ and both diagonal and non--diagonal elements of $\hat{W}$
are of the same order of magnitude. The SF then represents a
superposition of a narrow and a broad Lorentzian with widths
$W_{-}$ and $W_{+}$, respectively. This, in turn, gives rise to a short,
$\tau_{+}=W_{+}^{-1}=\tau/(1+q)$, and long, 
$\tau_{-}=W_{-}^{-1}=\tau/(1-q)$, decay times. In other words, the
state formed by two LS's, coupled via the continuum of 
states in the plane, is split into fast and slow decaying
components.\cite{sha94} 

The physical mechanism leading to the appearance of fast and slow
decaying components is, in fact, analogous to that of the Dicke effect
in the spontaneous emission of light by a gas.\cite{dicke} In
particular, the case of two LS's, coupled via the continuum
of electronic states with $\la_{_F}/r_{12}\gg 1$, is similar to the case
of a pair of atoms radiating a photon with the wave length $\la$ much
larger than interatomic distance $d$. For
$\la/d\gg 1$, which corresponds to the limit $q\rightarrow 1$,
the two atoms form a single quantum--mechanical system. The
electromagnetic field couples only to the symmetric state, 
which is the fast--radiating component, whereas for the
antisymmetric state (slow--radiating component) the corresponding
matrix element vanishes. The emission spectrum represents a narrow
peak with the width $W_{-}\rightarrow 0$ on top of a wide peak with
the width $W_{+}\rightarrow 2W$, where here $W^{-1}=\tau$ is the 
radiating time of an isolated atom. The wide peak, corresponding to the fast
time $\tau_{+}=\tau/2$, is a manifestation of the {\em superradiance},
that is {\em coherent} emission with the doubled rate, while the narrow
peak, corresponding to the {\em subradiance}, describes the 
{\em trapping of radiation} by atoms.\cite{voe96} Similarly, 
the first term of the SF (\ref{Spair1}), which turns into 
${1\over 2}S_0(\om)={1\over 2}\de(\om-\ep)$ for $r_{12}k_{_F}\ll 1$,
describes the {\em trapping of electronic states} by
the LS's (``subtunneling''), and the second term indicates that the
fast component decays into the continuum of states in the plane with
the doubled rate (``supertunneling'').

This analogy holds for an arbitrary number of LS's. The SF of $N$ LS's is
still given by Eqs.~(\ref{Spair}) (with factor $1/2$ replaced by
$1/N$) and (\ref{Gpair}), where $\hat{W}$ and $\hat{\ep}$ are now
$N\times N$ matrices. For identical LS's confined
within the area $\la_{_F}^2$, all the elements of $\hat{W}$ are again
of the same order of magnitude. To gain qualitative understanding, let
us assume them equal, $W_{ij}=qW$, with some $q\sim 1$. Then the 
generalization of Eq.~(\ref{Spair1}) reads

\be\label{SN}
S(\omega)={1\over \pi}
\Biggl[\left(1-{1\over N}\right){W_{-}\over (\omega-\ep)^2+W_{-}^2}
+{1\over N}{W_{s}\over (\omega-\ep)^2+W_{s}^2}\Biggr],
\ee
with $W_{s}=[1+q(N-1)]W$. We see again that as $q\rightarrow 1$, a fraction
$1-1/N$ of all states becomes trapped by the LS's, while the remaining
fraction $1/N$ is distributed in a wide interval $NW$. The latter
translates into the fast decay time $\tau_s=\tau/N$. This is again
completely analogous to the Dicke effect for $N$ atoms confined in a
volume with linear size much smaller than $\la$. 


With this understanding, let us turn back to our system of randomly
distributed LS's coupled to a 2D electron gas in strong
magnetic field. In a realistic system, in addition to LS's, a ``usual''
disorder is present in the electron plane, which we assume to be
uncorrelated. At the same time, the energy levels of LS's are not all
the same, but, in general, distributed within some interval. This
introduces into the problem yet another type of disorder, which is 
completely absent in the Dicke effect for a gas of 
{\em identical} atoms. As we will see below, the interplay of the two
types of disorder appears to be rather non--trivial.

An important parameter characterizing the system is the number of LS's
in the  area $\la_{_F}^2$. For the lowest LL, this parameter is just the 
``filling factor'' of LS's, 

\be\label{nu}
\nu=(2\pi l^2)n_{_{LS}},
\ee
where $n_{_{LS}}$ is LS's concentration and
$l$ is the magnetic length. For $\nu\ll 1$, the coupling between LS's
via the states in the plane is weak, and $S(\om)$ represents a
convolution of SF's of isolated LS's (coupled to the plane). In the
opposite limit, $\nu\gg 1$, the coupling between LS's is strong and,
as the above example suggests, nearly all electronic states should be
trapped by the LS's. Note however that in this example, the collective
(Dicke) state is characterized by the proximity of the parameter $q$
to unity: in the limit $q\rightarrow 1$, the fraction $1-1/N$ of
electronic states is trapped for {\em arbitrary} $N$ (which is
analogous to $\nu$). Although, it is not possible to introduce 
{\em a priori} a parameter similar to $q$ in a disordered 
system, one expects on physical grounds that with increasing
$\nu$ the system will find itself in the Dicke state. In particular,
for large $\nu$, one expects that the weight $1-1/\nu$ of the SF
$S(\om)$ will be carried by the ``bare'' SF, $S_0(\om)$, calculated in
the {\em absence} of tunneling. However, the interval of
$\nu$ within which the crossover between two regimes occurs, should
depend strongly on the disorder. As we show below, under certain
conditions the transition to the Dicke state can occur at some
{\em critical} value of $\nu$.

It is useful to view this system from a slightly different
angle. Namely, let us consider the effect of LS's on the electronic
states in the plane. Clearly, as the electron energy  approaches the LS
levels, the electron experiences {\em resonant scattering} by the LS's.
As a result, the electronic DOS should exhibit a sharp energy
dependence near the resonance. The character of this dependence can be
easily understood from the physical picture outlined above. Since with
increasing $\nu$, a larger fraction of the electronic states is
trapped by the LS's, the DOS should develop a {\em gap} in the limit
of large $\nu$. Incidentally, the fact that resonant scattering should 
lead to a gap for large LS's concentrations has long been known in
the 3D case for identical scatterers (in the absence of magnetic
field).\cite{iva,lif88,jau83} The above arguments suggest that
the resonant scattering and the Dicke effect are, in a certain sense, 
complementary to each other. The goal of the present paper is to
establish this relation in precise terms.

In fact, it is easy to see that the shape of the SF of LS's is
determined entirely by the resonant scattering. 
Indeed, the electron in the ``course'' of a single transition between
a LS and the plane is being scattered by the rest of LS's. Therefore,
the self--energy of a LS is simply proportional to the Green function
of 2D electron in the presence of  resonant scattering. This formal
relation indicates, however, that finding the SF, averaged 
over the positions and energies of LS's as well as over the in--plane
disorder, is, in general, a rather difficult task. In particular,
it requires the calculation of not only the averaged electron Green
function\cite{sha97} but, in fact, of all its moments. Nevertheless,
as we show below, for the lowest LL level the problem can be solved 
{\em exactly}. This solution, which is the main result of the present
paper, is possible due to the hidden supersymmetry of the lowest
LL.\cite{F,G}  

Let us briefly summarize our results. The exact expressions for the SF
$S(\om)$, and the DOS $g(\om)$, are multiparametric functions determined
by the LS filling factor $\nu$, the tunneling strength $\de$, the
Wegner's width $\Ga$ of the lowest LL (characterizing the in--plane
disorder), and the distribution function of the
LS levels $f_{\ga}(\ep-\bep)$, where $\bep$ is the average energy and
$\ga$ is the width. In the absence of coupling to the plane, $\de=0$,
the ``bare'' SF is simply $S_0(\om)=f_{\ga}(\om-\bep)$. In the
presence of coupling, $\de\neq 0$, we distinguish between two regimes
governed by the dimensionless parameter $\de^2/\ga\Ga$. 

In the weak coupling regime, $\de^2/\ga\Ga\ll 1$, we find that 
the transition to the Dicke state is smooth. In the limit of large
$\nu$, the fraction $1-1/\nu$ of electronic states is trapped by the
LS's, yielding $S(\om)=(1-1/\nu)S_0(\om)$. At the same time, the DOS
in the presence of resonant scattering exhibits a pronounced minimum
which develops into a gap with increasing $\nu$. The width of the gap
is independent of the disorder and is determined by the tunneling
strength and the LS concentration only. We demonstrate that this
behavior is {\em universal} and persists for arbitrary 
distribution of LS levels. 

In the strong coupling regime, $\de^2/\ga\Ga\gg 1$, the SF and the DOS
exhibit a rather complicated behavior. In the limit of vanishing
in--plane disorder and LS level spread, both $S(\om)$ and
$g(\om)$ are non--analytic in $\om$, turning to zero in a {\em finite}
energy interval. This gap originates from the infinite degeneracy of
the LL in the absence of disorder and is unrelated to the gap in the
weak coupling regime due to the trapping of states.

We find that in the case of strong coupling, the transition to the
Dicke state occurs at the {\em critical} concentration of LS's
corresponding to $\nu=1$: for {\em arbitrary} $\nu>1$, a 
fraction  $1-1/\nu$ of states is trapped by LS's. At the same time,
the DOS exhibits a seemingly similar behavior for $\nu<1$: a fraction
$1-\nu$ of states in the LL center remains unaffected by the resonant
scattering. The origin of such ``condensation of states''
is analogous to the one in the case of non--resonant
point--like scatterers.\cite{and74,B,G,J} For $\nu<1$, one can choose
as a basis linear combinations of unperturbed wave functions
vanishing at the positions of all LS's. This reduces the LL degeneracy by
a factor $1-\nu$, leaving this fraction of states unaffected. 

In fact, the similar behavior of $S(\om)$ for $\nu>1$ and
of $g(\om)$ for $\nu<1$ is not coincidental, but is a consequence of a
rather remarkable relation between the SF and the DOS in the absence
of disorder. We demonstrate that $S_(\om)$ and $g(\om)$ turn into each
other under the transformation $\nu\leftrightarrow 1/\nu$ and 
$\om\leftrightarrow \ep-\om$. This unexpected ``duality'' relates 
to each other the two phase transitions of entirely different physical
origins. 

The paper is organized as follows. In Section II we formulate the
model and derive the general expression for the SF. In Section III the
calculation of the averaged Green function of LS's is performed. The
analysis and numerical results are presented in Section IV. Section V
concludes the paper.

\section{Localized states and resonant scattering}

Consider a 2D electron gas in strong perpendicular magnetic field
in the presence of Gaussian random potential $V(\br)$ with correlator 

\be\label{corr}
\overline{V(\br)V(\br')}=w\de(\br-\br').
\ee
The electron plane is separated by a tunneling barrier from a plane of 
point--like LS's. We assume that the energy levels of LS's are
close to  the lowest LL and adopt the tunneling Hamiltonian

\be\label{th}
\hat{H}=\sum_{\mu}\ep_{\mu}a^{\dagger}_{\mu}a_{\mu}+
\sum_{i}\ep_{i}c^{\dagger}_{i}c_{i}+
\sum_{\mu,i}(t_{\mu i}a^{\dagger}_{\mu}c_{i}+\mbox{h.c.}).
\ee
Here $\ep_{\mu}$, $a^{\dagger}_{\mu}$ and $a_{\mu}$ 
are the eigenenergy, creation and annihilation operators of the 
eigenstate $|\mu\rangle$ of the Hamiltonian $H_0+V(\br)$, $H_0$ being
the Hamiltonian of free 2D electron in a magnetic field, 
$\ep_{i}$, $c^{\dagger}_{i}$ and 
$c_{i}$ are those for the $i$th localized state, and $t_{\mu i}$ 
is the tunneling matrix element. The latter is defined as

\bea\label{tun}
t_{\mu i}=\int d\br dz
\psi_{\mu}^{\ast}(\br,z)V_{i}(\br,z)\psi_{i}(\br,z)\simeq 
\psi_{\mu}^{\ast}(\br_i,z_i)
\int d\br dzV_{i}(\br,z)\psi_{i}(\br,z), 
\eea
where $V_{i}(\br,z)$ is the LS potential and
$\psi_{i}(\br,z)$ is its wave function.
In the perpendicular direction, the wave function 
$\psi_{\mu}^{\ast}(\br,z)$ decays as $e^{-\ka z}$, $\ka$ being the decay 
constant, while in the plane it behaves as eigenfunction 
$\psi_{\mu}^{\ast}(\br)$ of the Hamiltonian $H_0+V(\br)$. 
We assume that the tunneling barrier is high enough, so that the
dependence of $\ka$ on $\mu$ can be neglected.\cite{sha94} 
Thus, we have 

\be\label{tun1}
t_{\mu i}=\psi_{\mu}^{\ast}(\br_i,0)e^{-\ka z_i}
\int d\br dz V_{i}(\br,z)\psi_{i}(\br,z)
=\psi_{\mu}^{\ast}(\br_i)t_i,
\ee
with $t_i$ determined by the transparency of the barrier.

We are interested in the LS Green function, 

\be\label{Dformal}
D(\om)=\overline{N^{-1}\sum_i\langle i|(\om -\hat{H})^{-1}|i\rangle}
=\overline{N^{-1}\sum_i D_{i}(\om)},
\ee
where the overbar stands for averaging over positions and energies of LS's
as well as over the random potential $V$. Each term in the sum
(\ref{Dformal}) can be presented as  

\be\label{D}
D_{i}(\om)={1\over \om-\ep_i-\Si_i(\om)},
\ee
where $\Si_i(\om)$ is the self--energy resulting from a virtual
transition between the  $i$th LS and the plane. In the 
presence of several LS's, such  a transition includes also 
transitions between the plane and the rest of LS's. The latter
transitions result in coupling between LS's via the states in the
plane. Introducing the coupling matrix $\hat{T}$, 

\be\label{tran}
T_{ij}(\om)=\sum_{\mu}{t_{i\mu}t_{\mu j}\over \om-\ep_{\mu}},
\ee
the self-energy $\Si_i(\om)$ can be presented as

\bea\label{self}
\Si_i(\om)=T_{ii}+\sum_{j}{}\!' T_{ij}D_{0j}T_{ji}
+\sum_{jk}{}\!' T_{ij}D_{0j}T_{jk}D_{0k}T_{ki}+\cdots,
\eea
where $D_{0j}(\om)=(\om-\ep_j)^{-1}$, and the prime indicates that
the terms $j,k=i$ in the sums are omitted. 

It is convenient to recast $\Si_i(\om)$ in a different form. Using
Eq.~(\ref{tun1}), the coupling matrix can be written as   
 
\be\label{tran1}
T_{ij}(\om)=t_it_j\tilde{G}(\br_i,\br_j),
\ee
where $\tilde{G}(\br,\br^{\prime})
=\langle\br|(\om-H_0-V)^{-1}|\br^{\prime}\rangle$ is the Green
function of a 2D electron in the absence of LS's. After substituting
Eq.~(\ref{tran1}) into Eq.~(\ref{self}), the self--energy 
takes the form 

\bea\label{self2}
\Si_i(\om)=
&&
t_i^2\biggl[\tilde{G}(\br_i,\br_i)
+\int d\br \tilde{G}(\br_i,\br) U(\om,\br)\tilde{G}(\br,\br_i)
\nonumber\\
&&
+\int d\br d\br^{\prime} \tilde{G}(\br_i,\br)
U(\om,\br)\tilde{G}(\br,\br^{\prime})
U(\om,\br^{\prime})\tilde{G}(\br^{\prime},\br_i)
+\cdots\biggr],
\eea
with 

\be\label{res}
U(\om,\br)=\sum_j{t_j^2\over \om-\ep_j}\de(\br_j-\br).
\ee
The random potential $U(\om,\br)$ describes 
{\em resonant scattering} of electrons by LS's. It has a form similar to
that of the point--like scatterers. The crucial difference, however,
is that here the scattering strength depends on the proximity of the
electron energy to the LS levels. In particular, the potential
(\ref{res}) changes from repulsive to attractive as the electron
energy passes through the resonance. Since the LS positions
are random with uniform density $n_{_{LS}}$, the distribution function
of $U(\om)$ is Poissonian. Note that due to the spread in the LS
energies $\ep_j$ and tunneling amplitudes $t_j$, the 
scattering strengths in Eq.~(\ref{res}) are also random. 

Finally, after summation of the series (\ref{self2}), the
self-energy takes the compact form  

\be\label{self3}
\Si_i(\om)=t_i^2G(\br_i,\br_i),
\ee
where
\be\label{green}
G(\br,\br^{\prime})
=\langle\br|{1\over \om-H_0-V-U(\om)}|\br^{\prime}\rangle, 
\ee
is the Green function of 2D electron in the presence of resonant
scattering. 

In the following we assume that the magnetic field is strong and the 
scattering retains the electron in the lowest LL. While this condition
is standard for the white--noise potential, it seems to be more
restrictive for the resonant scattering. 
It should be noted, however, that the scattering strength is
effectively reduced by the spread in the LS levels. We also assume
that the tunneling barrier, separating LS's from the electron plane,
is high enough and neglect the difference between tunneling amplitudes
of different LS's,  setting $t_i=t$ in the rest of the paper.  

Equations (\ref{Dformal}),~(\ref{D}),~(\ref{self3}) and (\ref{green})
determine, in principle, the spectral function, 

\be\label{spe}
S(\om)=-{1\over\pi}\im D(\om)=-{1\over \pi}\im \overline{D_i(\om)}.
\ee
The averaged LS Green function, $D(\om)=\overline{D_i(\om)}$, can be
presented as the series

\bea\label{dav}
\overline{D_i(\om)}=\overline{1\over \om-\ep_i-t^2G(\br_i,\br_i)}
=\sum_{n=0}^{\infty}\left\langle 
{t^{2n}\over (\om-\ep)^{n+1}}\right\rangle_{\ep}G_n(\om),
\eea
where $G_n(\om)=\overline{G^n(\br_i,\br_i)}$ and 
$\langle\cdots\rangle_{\ep}$ denotes averaging over $\ep$. 
In obtaining Eq.~(\ref{dav}) we used the fact
that in Eqs.~(\ref{D}) and (\ref{self2}), the contribution of the 
$i$th LS into the potential (\ref{res}) is excluded.

The ensemble averaging in Eq.~(\ref{dav}) should be performed over
{\em both} random potentials $V$ and $U(\om)$. Calculation of 
$S(\om)$ requires then an averaging of not only the electron Green
function (\ref{green}), but, in fact, of all the moments $G_n(\om)$. 
Remarkably, for the lowest LL this averaging can be performed 
{\em exactly} by generalizing the approach of Ref.\onlinecite{G}

\section{Calculation of the spectral function}

In order to find
the moments $G_n(\om)$, we rely on the hidden supersymmetry of the
lowest LL.\cite{F,G} We start by noting that $G^n(\br,\br)$ can be
presented as a Gaussian functional integral over bosonic fields,

\bea\label{green1}
G^n(\br,\br)={(-i)^n\over n!}Z^{-1}\int\cD\vp\cD\vpb 
e^{iS}[\vp(\br)\vpb (\br)]^n,
\eea
with the action (here $\om_{+}=\om+i0$)

\bea
S[\vpb ,\vp]=\int d\br 
\vpb (\br)[\om_{+}-H_0-V-U(\om)]\vp(\br).
\eea
The normalization factor $Z^{-1}$ can be written as a fermionic integral,
$Z^{-1}=\int\cD\chi\cD\chib e^{iS}$, with the
same action $S[\chib,\chi]$. The fields $\vp(\br)$ and
$\chi(\br)$ are then projected onto the lowest LL subspace
according to (we measure all energies from the lowest LL)

\be\label{proj}
(\om-H_{0})\vp(\br)=\om\vp(\br),~~
(\om-H_{0})\chi(\br)=\om\chi(\br).
\ee
Choosing the symmetric gauge, this projection is achieved
with the representation

\bea\label{uv}
\vp =(2\pi l^2)^{-1/2}e^{-|z|^2/4l^2}u(z),~~
\chi=(2\pi l^2)^{-1/2}e^{-|z|^2/4l^2}v(z), 
\eea
where the bosonic field $u(z)$ and the fermionic field $v(z)$ are
analytic functions of the complex coordinate $z=x+iy$.
In terms of projected fields, Eq.~(\ref{green1}) takes the form  

\bea\label{green2}
G^n(\br,\br)={(-i)^n\over n!}{e^{-n|z|^2/2l^2}\over (2\pi l^2)^n}
\langle [u(z)\ub (\zc )]^n\rangle,
\eea
where angular brackets stand for the functional integration 
$\int\cD u \cD \ub \cD v \cD \vb e^{iS}$ with action

\be\label{act2}
S[u,v]=\int{d^{2}z\over 2\pi l^2}e^{-|z|^2/2l^2}(\ub u+\vb v)
[\om_{+}-V-U(\om)].
\ee
As a next step, one introduces Grassman (anticommuting) coordinates 
$\ta$ and $\tac$, satisfying 

\bea
\int d\ta = \int d\tac  =0, ~~
\int d\ta d\tac\tac\ta=\pi^{-1},
\eea
(normalized such that $\int d^2zd^2\ta e^{-|z|^2-\ta\tac}=1$), which together
with the coordinates $z$ and $z^{\ast}$ form the ``superspace''
$\xi=(z,\ta)$. One then defines the analytic ``superfields'' 

\bea\label{sup}
\Phi(z,\ta)&=&u(z)+\ta v(z)/\sqrt{2}l \, ,
\nonumber\\
\Phib (\zc,\tac)&=&\ub (\zc )+\tac \vb (\zc )/\sqrt{2}l \, , 
\eea
taking values in the ``superspace'' $\xi$.
Using the identities $\langle u\rangle=\langle v\rangle=0$, 
$\langle u\ub\rangle=\langle v\vb\rangle$, and 
$\langle (u\ub)^n\rangle=
n\langle u\ub \rangle\langle (u\ub)^{n-1}\rangle$, it is readily seen
that  the following chain of equalities holds 

\bea
\langle (\Phi\Phib)^n\rangle=\langle (u\ub)^n\rangle
+{n^2\ta\tac\over 2l^2}\langle v\vb\rangle\langle (u\ub)^{n-1}\rangle
=\left(1+{n\ta\tac\over 2l^2}\right)\langle(u\ub)^n\rangle
=e^{n\ta\tac/2l^2}\langle(u\ub)^n\rangle.
\eea
Thus, the correlation function (\ref{green2}) can be presented in
terms of a functional integral over superfields (\ref{sup}) as

\be\label{green3}
G^n(\br,\br)={(-i)^n\over n!}{e^{-n\xi\xic/2l^2}\over(2\pi l^2)^n}
\int\cD\Phi\cD\Phib e^{iS}[\Phi(\xi)\Phib (\xic)]^n, 
\ee
with $\xi\xic\equiv |z|^2+\ta\tac$. The action $S[\Phib,\Phi]$ in
Eq.~(\ref{green3}) is obtained from  Eq.~(\ref{act2}) by substituting

\bea
{e^{-|z|^2/2l^2}\over 2\pi l^2}(\ub u+\vb v)
=\int d^2\ta e^{-\xi\xic/2l^2}\Phib (\xic)\Phi(\xi)
\equiv Q(z,\zc).
\eea

We now perform the ensemble averaging over $V$ and $U(\om)$. The
Gaussian averaging of $\exp\left( - i\int V Q d^2 z\right)$ gives 
$\exp\left[-(w/2)\int Q^2 d^2z\right]$. The averaging over
the LS potential $U(\om)$, carried out with the Poissonian distribution
function,\cite{fri75} yields

\be\label{aver}
\overline{\exp\left( -i\int U Q d^2z\right)}
=\exp \left\{-n_{_{LS}}\int \left[1-
\left\langle \exp\left(-{it^2 Q\over \om-\ep}\right)
\right\rangle_{\ep}\right]d^2z\right\}. 
\ee
As a result, one obtains the following effective action

\bea\label{act3}
i\bar{S}[\Phi,\Phib]=&&i\om_{+}\int d^2\xi\, \al
-{\Ga^2\over 2}\int {d^{2}z\over 2\pi l^2}
\left(2\pi l^2\int d^2\ta \, \al \right)^2
\nonumber\\&&
-\nu\int {d^{2}z\over 2\pi l^2}\left\{1-\left\langle
\exp\left[
-{i\de^2\over \om-\ep}
\left(2\pi l^2\int d^2\ta \, \al\right)
\right]\right\rangle_{\ep}\right\},
\eea
with $d^2\xi=d^2zd^2\ta$ and 

\be\label{alpha}
\al(\xi,\xic)=e^{-\xi\xic/2l^2}\Phib(\xic)\Phi(\xi).
\ee
Here $\Ga=(w/2\pi l^2)^{1/2}$ is Wegner's width of the
lowest LL in the absence of LS's, $\nu=(2\pi l^2)n_{_{LS}}$ is
the ``filling factor'' of LS's, and the parameter 
$\de=t/(2\pi l^2)^{1/2}$ characterizes the tunneling. 

The action (\ref{act3}) possesses a supersymmetry, characteristic 
for the lowest LL.\cite{F,G} This symmetry between $z$ and $\ta$
coordinates is evident for the first term of Eq.~(\ref{act3}). It can
be made explicit for the second and the third terms also
by making use of the identity\cite{G} 

\bea\label{ident}
n\left(2\pi l^2 \int d^2\ta 
e^{-\ta\tac/2l^2} \Phib \Phi \right)^n
= 2\pi l^2 \int d^2\ta e^{-n\ta\tac/2l^2} (\Phib\Phi)^n.
\eea
This allows one to replace any functional of the form
$\int d^2z f\left(2\pi l^2\int d^2\ta\, \al\right)$ by the functional
$2\pi l^2\int d^2\xi h\left(\al\right)$ 
with $\partial h(x)/\partial x =f(x)/x$. The result is a 
manifestly supersymmetric action $\bar{S}=\int d^2\xi\cA(\al)$, where

\bea\label{act4}
i\cA(\al)=i\om_{+}\al -{\Ga^2\al^2\over 4}
-\nu\int_0^{\al}{d\beta\over\beta}\left[1-
\left\langle\exp\left(-{i\de^2 \beta\over \om-\ep}\right)
\right\rangle_{\ep}\right].
\eea
With all three terms now depending on superfields via 
$\al(\xi,\xic)$ only, the perturbation series (with respect to the
second and third terms) for the moments 
$\langle [\Phi(\xi)\Phib(\xic)]^n\rangle$
drastically  simplifies. One notices that transformations of the form 

\bea\label{trans}
\Phi(\xi)\rightarrow \Phi(\xi-\eta)e^{\xi\etac/2l^2-\eta\etac/4l^2},
\eea
generate translations of $\al$ in the superspace, 
$\al(\xi,\xic)\rightarrow\al(\xi-\eta,\xic-\etac)$, and hence leave
the action $\bar{S}=\int d^2\xi\cA(\al)$ invariant. This leads to 

\bea\label{cn}
\langle [\Phi(\xi)\Phib(\etac)]^n\rangle=C_ne^{n\xi\etac/2l^2},
\eea
and from Eq.~(\ref{green3})

\bea\label{mom}
G_n(\om)={(-i)^nC_n\over n!(2\pi l^2)^n},
\eea
where the coefficients $C_n$ are $\xi$--independent. For free electrons
one has $C_n=i^nn!/\om_{+}^n$. With the action (\ref{act4}), the
coefficient $C_1$ determines the averaged electron Green function,
$G(\om)\equiv G_1(\om)$, in the presence of resonant
scattering.\cite{sha97}

For an arbitrary $n$, the moments 
$\langle [\Phi(\xi)\Phib(\xic)]^n\rangle$ can be derived 
exactly by extending the arguments of Ref.\onlinecite{G} to the case
of $n$--point correlators.
A diagram with $N$ internal lines contains $N$ free propagators of the
form $-i\langle\Phi(\xi)\Phib(\etac)\rangle_0
=e^{\xi\eta^{\ast}/2l^2}/\om_{+}$, while the contribution of each
vertex is proportional to 
$e^{-m\zeta\zeta^{\ast}/2l^2}$, $2m$ being the number of lines
entering the vertex. After extracting a common factor 
$e^{n\xi\xic/2l^2}$, in accordance with Eq.~(\ref{cn}), the
contribution of a diagram can be written as $c_NK_N$, where $K_N$ is a
($N$--fold) Gaussian integral in superspace.
The value of the latter is unity due to the exact cancellation between
$z$ and $\ta$ integrals. The remaining coefficients $c_N$ can be
generated within the zero--dimensional field theory with partition
function $Z_0=\int d^2\phi e^{i\cA(\phi\phic)}$, where $\phi$ is a
complex number and the action $\cA(\phi\phic)$ is the same as in
Eq.~(\ref{act4}). The coefficients $C_n$ are then found as ratios of
two ordinary integrals 
 
\bea\label{Cn}
C_n=Z_0^{-1}\int d^2\phi e^{i\cA (\phi\phic)}(\phi\phic)^n. 
\eea
With such $C_n$ and with help of Eqs.~(\ref{mom}) and (\ref{dav}),
we finally arrive at the following {\em exact} expression for the
Green function of LS's

\bea\label{dav1}
D(\om)&=&{\pi\over Z_0}
\int_0^{\infty}d\al e^{i\cA(\al)}
\left\langle{1 \over \om-\ep}
\exp\left(
-{i\de^2 \al\over \om-\ep}
\right)\right\rangle_{\ep},
\nonumber\\
Z_0&=&\pi \int_0^{\infty}d\al e^{i\cA(\al)},
\eea
with $\cA(\al)$ given by Eq.~(\ref{act4}). 

It is also useful to present
this expression in a different form. To do this, we  introduce the
distribution function of LS energies, $f_{\ga}(\ep-\bep)$, where
$\bep$ is the average energy and $\ga$ is the width. It can be easily
seen from Eq.~(\ref{act4}) that 

\be\label{integ}
\left\langle{1 \over \om-\ep}\exp\left(-
{i\de^2 \al\over \om-\ep}\right)\right\rangle_{\ep}
=D_0(\om)+{i\over \nu}{\partial \cA(\al)\over \partial\bep},
\ee
where

\be\label{D0}
D_0(\om)\equiv \langle D_{0j}(\om)\rangle_{\ep}
=\int d\ep {f_{\ga}(\ep)\over \om-\bep-\ep}
\ee
is the averaged LS Green function in the absence of coupling to the
electron plane. Combining Eqs.~(\ref{dav1}) and (\ref{integ}), we
finally obtain 

\be\label{dav2}
D(\om)=D_0(\om)+{1\over \nu}{\partial \ln Z_0 \over \partial\bep}.
\ee
The analysis of this expression will be performed in the following section.

\section{Discussion and numerical results}

The final expression for the SF, Eq.~(\ref{spe}), appears to be rather
involved and its analysis requires distinguishing between several
cases. It is convenient to perform explicitly the averaging over $\ep$
in the action (\ref{act4}). The result reads\cite{remark1}

\bea\label{exp}
i\cA(\al)=i\om_{+}\al -{\Ga^2\al^2\over 4}
-\nu\int_0^{\infty}{dx\over x}\tilde{f}_{\ga}(x)e^{i(\om-\bep)x}
\left[1-J_0\left(2\de\sqrt{x\al}\right)\right],
\eea
where $\tilde{f}_{\ga}(x)$ is the Fourier transform of the
distribution function $f_{\ga}(\ep)$. With such  $\cA(\al)$,
one can obtain from Eq.~(\ref{dav2})

\be\label{dav3}
D(\om)=-i\int_0^{\infty}dx\tilde{f}_{\ga}(x)e^{i(\om-\bep)x}
\left\langle J_0\left(2\de\sqrt{x\al}\right)\right\rangle_{\al},
\ee 
where $\langle\cdots\rangle_{\al}$ stands for the average with the
partition function $Z_0$ from Eq.~(\ref{dav1}). The electron Green
function, $G(\om)$, is given by a slightly simpler expression

\be\label{egreen}
(2\pi l^2)G(\om)=-i\langle\al\rangle_{\al},
\ee
which follows from Eqs.~(\ref{mom}) and (\ref{Cn}) with $n=1$. The
form (\ref{exp}) of the action $\cA(\al)$ introduces the dimensionless
parameter $\de^2/\ga\Ga$ which represents the relative strength of the
coupling between LS's and electronic states in the plane. 

\subsection{Weak coupling}

Consider first the case of weak coupling, $\de^2/\ga\Ga\ll 1$. 
Substituting 

\bea
\left\langle J_0\left(2\de\sqrt{x\al}\right)\right\rangle_{\al}
&&
=\exp \ln \left\{1-\left[1-\left\langle J_0\left(2\de\sqrt{x\al}\right)
\right\rangle_{\al}\right]\right\}
\nonumber\\
&&
\simeq \exp(-\de^2 \left\langle\al\right\rangle_{\al}x),
\eea
into the rhs of Eq.~(\ref{dav3}), we obtain 

\be\label{dav4}
D(\om)=\left\langle {1\over \om-\ep-\Si(\om)}\right\rangle_{\ep},
\ee
with

\be\label{self4}
\Si(\om)=-i\de^2\langle\al\rangle_{\al}=\de^2(2\pi l^2)G(\om).
\ee 
Thus, the self-energy in this case is proportional to the averaged
electron Green function in the presence of resonant scattering. 
In particular, the width of the SF is determined by the
electronic DOS, 

\be
g(\om)=-{1\over\pi}\im G(\om).
\ee
Note that Eq.~(\ref{dav4}) could be also 
readily obtained from Eq.~(\ref{D}) by substituting the averaged
$\overline{\Si_i(\om)}=\Si (\om)$ from Eq.~(\ref{self3}). 

To simplify the analysis, let us assume that the distribution of LS
levels is Lorentzian (numerical calculations below are performed
with the more realistic Gaussian distribution\cite{med97}). Then the
averaging in Eq.~(\ref{dav4}) can be done analytically, yielding 

\be\label{specweak}
S(\om)={1\over \pi}{\ga +\ga_1 \over (\om-\bep-\ep_1)^2+(\ga+\ga_1)^2},
\ee
with 

\bea\label{epga1}
\ep_1(\om)&=&\de^2(2\pi l^2)\re G(\om),
\nonumber\\ 
\ga_1(\om)&=&-\de^2(2\pi l^2)\im G(\om)=\pi\de^2(2\pi l^2)g(\om).
\eea
Consider first the case of isolated LS, that is $\nu=0$ in
Eq.~(\ref{exp}). Then we have $G(\om)=G_W(\om)$, where $G_W(\om)$ is
Wegner's Green function.\cite{F} This leads to

\bea\label{ep1ga1}
\ep_1&=&{2\de^2\over\Ga}\left[{\om\over\Ga}
-{2\over\pi}
\frac{e^{\om^2/\Ga^2}\int_0^{\om/\Ga}dte^{t^2}}
{1+\left({2\over\sqrt{\pi}}\int_0^{\om/\Ga}dte^{t^2}\right)^2}\right],\\
{}\nonumber\\
\ga_1&=&{2\de^2\over\sqrt{\pi}\Ga}\frac{e^{\om^2/\Ga^2}}
{1+\left({2\over\sqrt{\pi}}\int_0^{\om/\Ga}dte^{t^2}\right)^2}.
\eea
The renormalization of the LS energy, $\ep_1$, is a slow function of
$\om$: $\ep_1\simeq 2\de^2\om(\pi-2)/\pi \Ga^2$ for $\om\ll\Ga$ and 
$\ep_1\simeq \de^2/\om$ for $\om\gg\Ga$, 
so that its role is relatively unimportant.  In contrast, the
renormalization of the width, $\ga_1$, 
being proportional to the DOS, is a sharp function of $\om$: 
$\ga_1=2\de^2/\sqrt{\pi}\Ga$ for $\om\ll\Ga$
and $\ga_1=(2\de^2\sqrt{\pi}\om^2/\Ga^3)e^{-\om^2/\Ga^2}$ for
$\om\gg\Ga$. Therefore, for $\bep\lesssim \Ga$, the effective width, 
$\ga+\ga_1$, experiences a
sharp increase as $\om$ approaches $\bep$. This, in turn, leads to a
minimum in the SF (see Fig. 1).

Let us now turn to the case of many LS's ($\nu\neq 0$). 
The Bessel function in Eq.~(\ref{exp}) can be expanded to first
order in $\de^2/\ga\Ga$, yielding 
 
\be\label{G1}
G(\om)=G_W\left(\om-\nu\de^2D_0\right),
\ee
with $D_0(\om)$ given by Eq.~(\ref{D0}). 
Thus, in this case the energy of the electron is shifted by an
amount proportional to the {\em average} potential
$\langle U(\om,\br)\rangle_{\ep,\br}=\nu\de^2D_0(\om)$. At the same
time, the fluctuations of $U(\om,\br)$, which are described by terms
of higher orders in $\de^2/\ga\Ga$, are small. In other words, the
effect of scattering of the electron by LS's is reduced to that of an
{\em effective medium}. For the usual, non--resonant scattering 
($\om$--independent $U$), this would merely result in renormalization
of the energy by a constant. In the case of resonant scattering,
however, the average potential is a {\em complex} quantity. 
Its imaginary part, 
which originates from the spread in the LS levels, is a sharp function
of $\om$. This affects strongly the shape of the DOS and, in turn, of
the SF.

For a low LS concentration, $\nu\de^2/\ga\Ga\ll 1$, and for
$\om\sim\bep\ll\Ga$, we find that the change in the DOS, $\de g(\om)$,
is given by 

\be\label{dg}
{\de g \over g}={\de \ga_1 \over \ga_1}
\simeq -{\pi-2\over\sqrt{\pi}}{\nu\de^2\pi\over \Ga}S_0(\om),
\ee
where $S_0(\om)=\pi^{-1}\im D_0(\om)$ is the ``bare'' SF in the
absence of coupling to the plane. We see that the DOS in the presence
of resonant scattering exhibits a minimum. This, in turn, leads to a 
maximum in the SF via the renormalization of its width,
$\ga_1$. Numerical results for several sets of parameters are shown in
Figs.~1 and 2. 

With increasing LS concentration, the SF develops a pronounced peak
which saturates for large $\nu$ (see Fig.~3). In order to understand
this behavior, let us consider the case when the LS concentration is
high, $\nu\gg 1$, so that $\nu\de^2/\ga\Ga\gg 1$, but at the same time
the scattering remains weak. Then, the fluctuations of the random
potential $U(\om)$ are still suppressed, but the argument of $G_W$ in
Eq.~(\ref{G1}) is large. Presenting $G_W$ as [see Eq.~(\ref{egreen})]

\be\label{GW}
G_W\left(\om-\nu\de^2D_0\right)
=-{1\over 2\pi l^2}{\partial\over\partial\om_{+}}
\ln\int_0^{\infty}d\al\exp\left[i\al\left(\om_{+}-\nu\de^2D_0\right)
-{\al^2\Ga^2\over 4}\right],
\ee
we notice that for $\om-\bep\sim \ga$ one has $\im D_0\sim 1/\ga$, 
so that the last term in the exponent can be omitted. This gives

\be\label{GW1}
G_W\left(\om-\nu\de^2D_0\right)\simeq {1\over 2\pi l^2}
{1\over \om-\nu\de^2D_0(\om)}.
\ee
If $\bep$ is not in the LL tail, the first term in the denominator
can be neglected. For the Lorentzian distribution,
$D_0(\om)=(\om-\bep+i\ga)^{-1}$, this readily leads to
$\ep_1=(\bep-\om)/\nu$ and $\ga_1=\ga/\nu$, and we obtain

\be\label{univ}
S(\om)={1\over \pi (1+\nu^{-1})}{\ga\over(\om-\bep)^2+\ga^2}
\simeq\left(1-{1\over\nu}\right)S_0(\om).
\ee
We see that for large LS concentration, almost the entire weight of
$S(\om)$ is carried by the ``bare'' SF, $S_0(\om)$, meaning  
that the fraction $1-1/\nu$ of electronic states is trapped
by LS's. The remaining $1/\nu$ fraction of states is carried by the
tails, $\om-\bep\gg \ga$, which become longer as $\nu$ increases (see
Fig.~3). As was discussed in the Introduction, such form of $S(\om)$
is a manifestation of the Dicke effect. Clearly, in the case of weak
coupling, the transition to the Dicke state is smooth.

Although Eq.~(\ref{univ}) was derived  for the Lorentzian form of
$f_{\ga}(\ep)$, this behavior persists for an
arbitrary distribution of LS levels. Indeed, for $\bep\lesssim \Ga$
and $\om-\bep\sim \ga$, Eqs.~(\ref{dav4}), (\ref{self4}), (\ref{G1}),
and (\ref{GW1}) yield 

\be\label{Duniv}
D(\om)=D_0[\om+1/\nu D_0(\om)].
\ee
Since for such $\om$ one has $D_0(\om)\sim 1/\ga$, we see that $S(\om)$
is again given by $S_0(\om)$ up to a small fraction $1/\nu$.

At the same time, with increasing $\nu$ the DOS exhibits a pronounced
minimum which turns into a {\em gap} as $\nu$ becomes large (see Fig.~4). 
For $\nu\gg 1$, the width of the gap is of the order of the separation
between peaks. The latter can be estimated from the condition that
the argument of the Green function (\ref{GW}) turns to zero. Since
$D_0(\om)\simeq 1/\om$ for large $\om$, one easily finds that this
separation is $2\sqrt{\nu}\de$. Thus, the width of the gap is
universal and independent of the disorder. Within the gap, the DOS can
be estimated from Eq.~(\ref{GW1}) as $(2\pi l^2)g(\om)\sim \ga/\nu\de^2$. 

\subsection{Strong coupling}

In the case of strong coupling, $\de^2/\ga\Ga\gg 1$,
Eqs.~(\ref{dav4}) and (\ref{self4}) do not apply and,
in general, all moments $G_n$ contribute to $D(\om)$. 
Nevertheless, the SF and the DOS appear to be
intimately related. In order to reveal this relation, let us consider
the limit of vanishing disorder, with $\Ga/\de\ll 1$ and $\ga/\de\ll 1$. In
this case the second term in the action (\ref{act4}) can be omitted
and no energy averaging is implied. Then it is easy to see that in the
energy interval $\om(\ep-\om)>0$, the integration path in the
$\al$--integral for the partition function, 
$Z_0=\pi\int_0^{\infty}d\al e^{i\cA(\al)}$, can be 
rotated by $e^{-i\pi{\rm sgn}(\om-\ep)/2}$, resulting in purely real
$i\cA$. After rescaling the integration variable $\al$, the partition
function takes the form 

\bea\label{Z0}
Z_0={\pi(\om-\ep)\over i\de^2}\int_0^{\infty}d\al
\exp\left[ -\al{\om(\ep-\om)\over\de^2}
-\nu\int_0^{\al}{d\beta\over\beta}
\left(1-e^{-\beta}\right)\right].
\eea
With this $Z_0$, one obtains from Eq.~(\ref{dav2}) after some
algebra\cite{remark2} 

\bea\label{dav5}
D(\om)=\left(1-{1\over \nu}\right){1\over \om-\ep}
+{1\over \nu}{\partial\over\partial\ep}
\ln\int_0^{\infty}{d\al\over\al^{\nu}}
\exp\left[ -\al{\om(\ep-\om)\over\de^2}
-\nu\int_{\al}^{\infty}{d\beta\over\beta}e^{-\beta}
\right].
\eea
The second term can be written as  
$-(\om/\de^2\nu)\langle\al\rangle_{\al}$ [calculated with
the partition function (\ref{Z0})]. On the other hand, the same
manipulations with the electron Green function (\ref{egreen}) give 
$G(\om)=[(\ep-\om)/\de^22\pi l^2]\langle\al\rangle_{\al}$. This leads 
to the following relation 

\be\label{rel}
\nu(\ep-\om)D(\om) + \om (2\pi l^2) G(\om)=1-\nu.
\ee
The fact that $\im\ln Z_0$ has no energy dependence for
$\om(\ep-\om)>0$ implies that both the SF and the DOS
should exhibit a gap in this energy interval (see Fig.~5). However, in
contrast to the weak coupling case, here the gap is not related to 
the Dicke effect. The physical origin of this gap can be understood
from the following reasoning.\cite{sha97} In the absence of in--plane
disorder (small $\Ga$), the LL broadening comes from the resonant
scattering alone. Then the scattering potential (\ref{res}) appears to
be attractive for $\om<\ep$, pulling  the states from the LL center to
the {\em left}, while for $\om>\ep$ it is repulsive, pushing the
states to the {\em right}. At the same time, in the absence of LS
level spread (small $\ga$), the finite width of $S(\om)$ comes from
transitions between the LS's and the electron plane. Therefore, the
absence of states in the latter leaves $S(\om)$ unaffected [that is
$S(\om)=0$] in the same energy interval. It should be emphasized that
$S(\om)$ and $g(\om)$ are {\em non--analytic}, turning to zero for
arbitrary $\nu$ (in the weak coupling case, $S(\om)$ and $g(\om)$ are
finite for all $\om$, vanishing only in the limit
$\nu\rightarrow\infty$).  

Near the gap edges, $\om(\om-\ep)\rightarrow 0_{+}$, the behavior of
$S(\om)$ and $g(\om)$ depends strongly on the value of
$\nu$. The integral in Eq.~(\ref{dav5}) is similar to that already
analyzed in Ref.\onlinecite{G}. Consider first the case
$\nu<1$. The second term in Eq.~(\ref{dav5}) can be split as 

\bea\label{term}
&&
{1\over \nu}{\partial\over\partial\ep}\Biggl\{
(\nu-1)\ln[\om(\ep-\om)]
\nonumber\\
&&
+\ln\left[1-{1\over \Ga(1-\nu)}
\left[{\om(\ep-\om)\over\de^2}\right]^{1-\nu}
\int_0^{\infty}{d\al\over\al^{\nu}}
e^{-\al\om(\ep-\om)/\de^2}
\left[1-
\exp\left(
-\nu\int_{\al}^{\infty}{d\beta\over\beta}e^{-\beta}
\right)\right]\right]\Biggr\},
\eea
where $\Ga(x)$ is the Gamma--function. The derivative of the first
term in the braces cancels the first term 
in Eq.~(\ref{dav5}). The second term, after analytical continuation 
$\om(\ep-\om)\rightarrow e^{-i\pi}\om(\om-\ep)$, gives

\be\label{Sgap}
S(\om)={A\over \nu}{1\over|\om-\ep|^{\nu}}
\left|{\om\over \de^2}\right|^{1-\nu},
\ee
where $A$ is a $\nu$--dependent constant\cite{remark3} ($A\sim\nu^2$ as
$\nu\rightarrow 0$). 

The SF diverges at one edge of the gap,
$\om\rightarrow\ep$, and is continuous at the other, 
$\om\rightarrow 0$. The exactly opposite behavior is exhibited by the
DOS, for which we obtain from  Eq.~(\ref{rel})

\bea\label{dos}
(2\pi l^2)g(\om)=(1-\nu)\de(\om)+
\nu\, {\om-\ep \over \om}\, S(\om).
\eea
Aside from the first term, the behavior of the DOS near the gap edges,
$g(\om)\propto |\om|^{-\nu}|\om-\ep|^{1-\nu}$, ``mirrors'' that of the SF.
Figure 5 shows that the similarity is striking over the entire energy
range.

The first term in Eq.~(\ref {dos}) indicates that a fraction $1-\nu$
of states remains unaffected by the resonant scattering. Such a
``condensation of states'' originates from the residual LL degeneracy
left after arranging the unperturbed wave--functions to vanish
at the positions of all LS's, and is similar to that in the case
of point--like scatterers with constant scattering
strength.\cite{and74,B,G,J} In fact, the analogy extends also to the
intricate structure of the DOS away from the gap. In particular, the
smaller peaks correspond to the singularities\cite{G} in $g(\om)$ at
integer values of $\om(\om-\ep)/\de^2$; with increasing $\ga$ they are
washed out. Similar structure appears also in $S(\om)$; here it is
washed out with increasing $\Ga$ (see Fig.~5). 

In the case of $\nu\geq 1$, the DOS at  
$\om(\om-\ep)\rightarrow 0_{+}$ can be found in a similar
manner. The result reads\cite{remark3}

\bea\label{ggap}
(2\pi l^2)g(\om)&\propto& {1\over |\om|}
\ln^{-2}\left[{\om(\om-\ep)\over\de^2}\right],
~~~\mbox{for}~~\nu = 1,
\nonumber\\
&\propto&{|\om|^{\nu-2}}
\left|{\om-\ep\over \de^2}\right|^{\nu-1},
~~~\mbox{for}~~\nu > 1.
\eea
Note that for $\nu\geq 1$, the LL degeneracy is lifted completely, so
that no condensation of states occurs. Instead,  according to
Eq.~(\ref{rel}), the SF represents a sum of two terms 

\be\label{spectral}
S(\om)=\left(1-{1\over \nu}\right)\de(\om-\ep)
+{1\over \nu}\, {\om\over \om-\ep}\, (2\pi l^2)g(\om).
\ee
Since in the absence of disorder $S_0(\om)=\de(\om-\ep)$, we observe
again that the fraction $1-1/\nu$ of all states is trapped 
by LS's, while the tails of $S(\om)$, given by the second term, are
suppressed by the factor $1/\nu$. However, in contrast to the weak
coupling case, here such behavior persists not only for large, but for
{\em arbitrary} filling factor $\nu > 1$. It is instructive to
compare SF from Eq.~(\ref{spectral}) to the SF from 
Eq.~(\ref{SN}). In the latter case, $N$ identical LS's, confined to
the area $\la_{_F}^2$, form a coherent (Dicke) state in the limit
$q\rightarrow 1$. In former case, the transition into the Dicke state
occurs at {\em critical} filling factor $\nu=1$.   

Thus, in the absence of disorder, the system undergoes two
types of transitions at $\nu=1$: the condensation of states in the LL
center for $\nu<1$, and the trapping of states by LS's for $\nu>1$. It
should be emphasized that the two phase transitions have entirely
different physical origins and exist independently of each other. The
former is caused by the LL degeneracy and persists also for non--resonant
scattering; the latter, being a manifestation of the Dicke effect,
takes place also in the absence of magnetic field. Nevertheless, the one
can be derived from the other due to a surprising ``duality'' relation.
Namely, it is readily seen from Eqs.~(\ref {dos}) and 
(\ref {spectral}) that the SF and the DOS turn into each other,
$(2\pi l^2)g(\om)\leftrightarrow S(\om)$, under the transformation

\be\label{dual}
\nu\leftrightarrow 1/\nu,~~~\om\leftrightarrow \ep-\om.
\ee
It is rather remarkable that the resonant scattering and the Dicke
effect can be unified in such simple manner.

\section{conclusion}

We have shown that a system of LS's coupled to the 2D electron gas in
strong magnetic field exhibits a kind of collective behavior similar
to the Dicke effect. For high enough LS concentrations, the trapping
of electronic states by the LS's takes place, which is analogous to
the Dicke subradiance. Such trapping appears to be complementary to
the gap in the DOS in the presence of resonant scattering. 

Although our derivation was restricted to the lowest LL, we believe
that our results are more general and remain valid for higher LL's.
There is little doubt that the gap in the DOS is a rather general
feature. A much more subtle question is related to the type of the
transition to the Dicke state. It seems obvious that in a disordered
system this transition should be smooth. In a clean system, we have
shown that this is, in fact, a phase transition. However, this result
appears to be specific to the system in magnetic field, as indicated
by the existence of the duality between the trapping of states by LS's
and the condensation of states in the LL center. The latter
transition, being caused by the LL degeneracy, takes place for all LL
numbers\cite{J} (as far as LL mixing is neglected). Therefore, we
believe that for higher LL's, the transition to the Dicke state should
also occur at critical filling factor $\nu=1$, although we have not
proved the duality relation in the general case. In the absence of
magnetic field, however, the question remains open. 

As a possible experimental realization, we suggest a system of
self-assembled quantum dots separated from a 2D electron gas by a
tunable tunneling barrier. Due to the ultra-narrow distribution of
dots' sizes, the spread in their energy levels, $\ga$, does not
exceed\cite{med97} 10.0 meV. Although, it is hard to achieve 
the inter--dot separation much smaller than the Fermi
wave--length, a condition $\nu\sim 1$ seems to be quite resonable. 
For a considerable resonant scattering effect, one has to have 
$\de^2/\ga\Ga\sim 1$. For a typical LL width $\Ga\sim 1$ meV, this
condition implies that the tunneling parameter $\de$ should be about
several meV. We believe that the significant drop in the mobility,
observed (at zero field) by authors of Ref.\onlinecite{sak95}, should
be attributed to the gap in the DOS.

\acknowledgments
Illuminating discussions with M. E. Raikh are gratefully acknowledged.
This work was supported in part by US Department of Energy Grant
No. DE-FG02-91ER45334.

\begin{figure}
\label{fig:1}
\caption{
The SF at (a) $\nu=0$ (isolated LS), and 
(b) $\nu=1.5$, with  $\Ga/\ga=1.0$ and $\bep=0$, is shown 
for $\de/\ga=0.1$ (dot--dashed line), $\de/\ga=0.5$ 
(long--dashed line), $\de/\ga=1.0$ (dashed line), 
$\de/\ga=1.5$ (dotted line), and $\de/\ga=2.5$ (solid line).
}
\end{figure}

\begin{figure}
\label{fig:2}
\caption{
The DOS [in units of $g_1=(2\pi l^2)^{-1}\Ga^{-1}$] for
a strong in-plane disorder, $\de/\Ga=0.3$, with $\bep=0$ and $\nu=1.5$,
is shown for $\ga/\de=0.1$ (solid line), $\ga/\de=0.5$ (dotted
line), $\ga/\de=1.0$ (dashed line), $\ga/\de=2.0$ (long--dashed
line), and $\ga/\de=10.0$ (dot--dashed line).
}
\end{figure}

\begin{figure}
\label{fig:3}
\caption{
(a) The SF with  $\Ga/\ga=1.0$, $\de/\ga=1.0$, and $\bep=0$
is shown for $\nu=0$ (dot--dashed line), $\nu=0.8$ 
(long--dashed line),  $\nu=3.2$ (dashed line), $\nu=10.0$ 
(dotted line), and $\nu=16.0$ (solid line).
(b) The SF with $\Ga/\ga=2.0$, $\de/\ga=2.0$, and $\bep=0$
is shown for $\nu=0$ (dot--dashed line), $\nu=0.4$ (long--dashed
line), $\nu=1.6$ (dashed line), $\nu=6.0$ (dotted
line), and $\nu=16.0$ (solid line).
}
\end{figure}

\begin{figure}
\label{fig:4}
\caption{
(a) The DOS with $\ga/\Ga=1.0$, $\de/\Ga=1.0$, and $\bep=0$
is shown for $\nu=0$ (dot--dashed line), $\nu=0.8$ 
(long--dashed line),  $\nu=3.2$ (dashed line), $\nu=10.0$ 
(dotted line), and $\nu=16.0$ (solid line).
(b) The DOS with $\ga/\Ga=0.5$, $\de/\Ga=1.0$, and $\bep=0$
is shown for $\nu=0$ (dot--dashed line), $\nu=0.4$ 
(long--dashed line), $\nu=1.6$ (dashed line), $\nu=6.0$ 
(dotted line), and $\nu=16.0$ (solid line).
}
\end{figure}

\begin{figure}
\label{fig:5}
\caption{
(a) The SF for a small LS level spread, $\ga/\de=0.1$, 
with $\nu=0.8$ and $\bep/\de=1.0$ is shown for 
$\Ga/\de=0.1$ (solid line), $\Ga/\de=0.3$ (dotted line),
$\Ga/\de=0.5$ (dashed line), and $\Ga/\de=1.0$ (long--dashed line). 
(b) The DOS [in units of $g_2=(2\pi l^2)^{-1}\de^{-1}$] for a weak
in-plane disorder, $\Ga/\de=0.1$, with  
$\bep/\de=1.0$ and $\nu=0.8$, is shown for $\ga/\de=0.1$, (solid line), 
$\ga/\de=0.3$ (dotted line), $\ga/\de=0.5$ (dashed line), and
$\ga/\de=1.0$ (long--dashed line).
}
\end{figure}

\begin{figure}[htb]
\epsfxsize=6.5in
\epsfbox{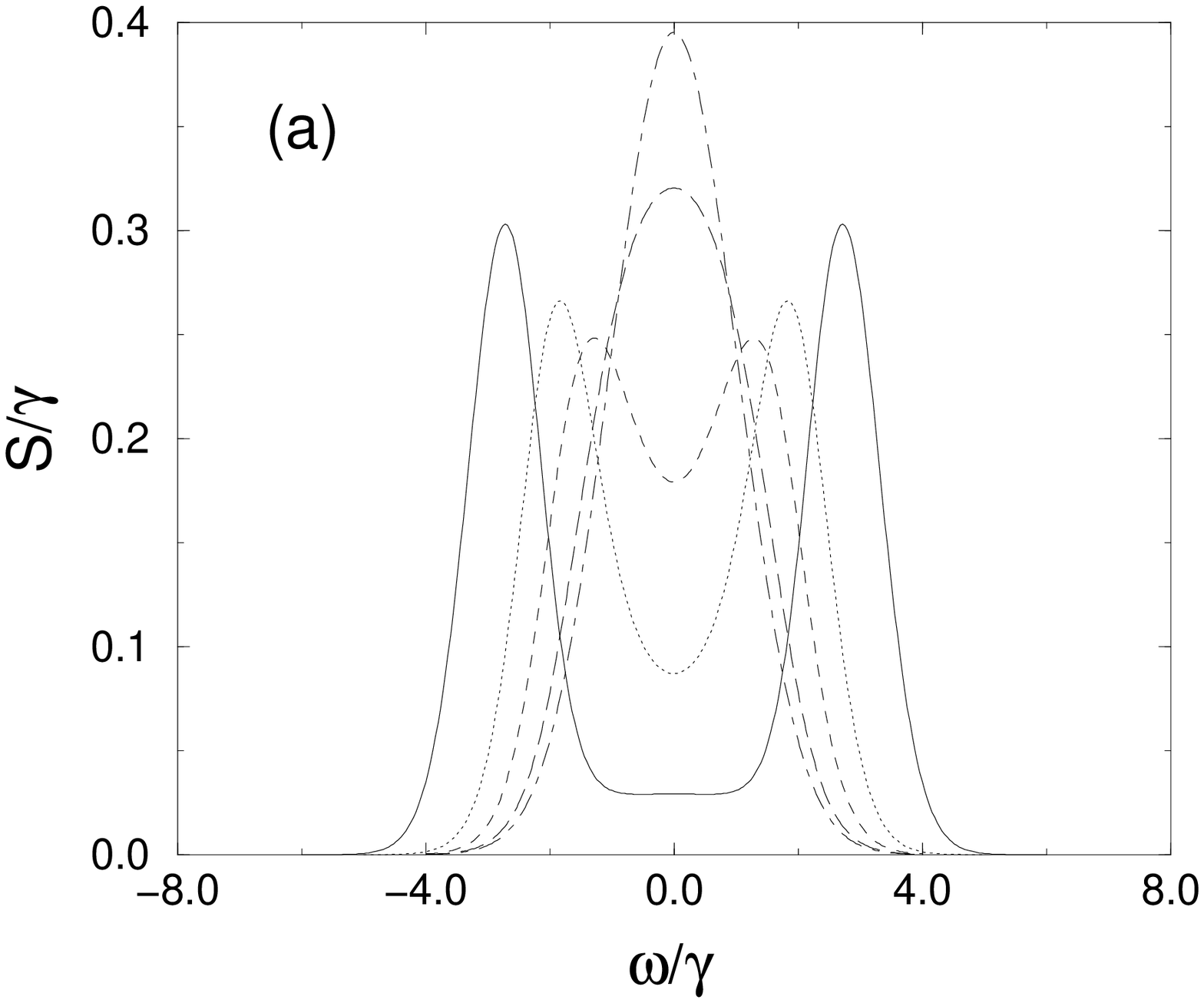}
\end{figure}

\vspace{70mm}

\centerline{FIG. 1}

\clearpage

\begin{figure}[htb]
\epsfxsize=6.5in
\epsfbox{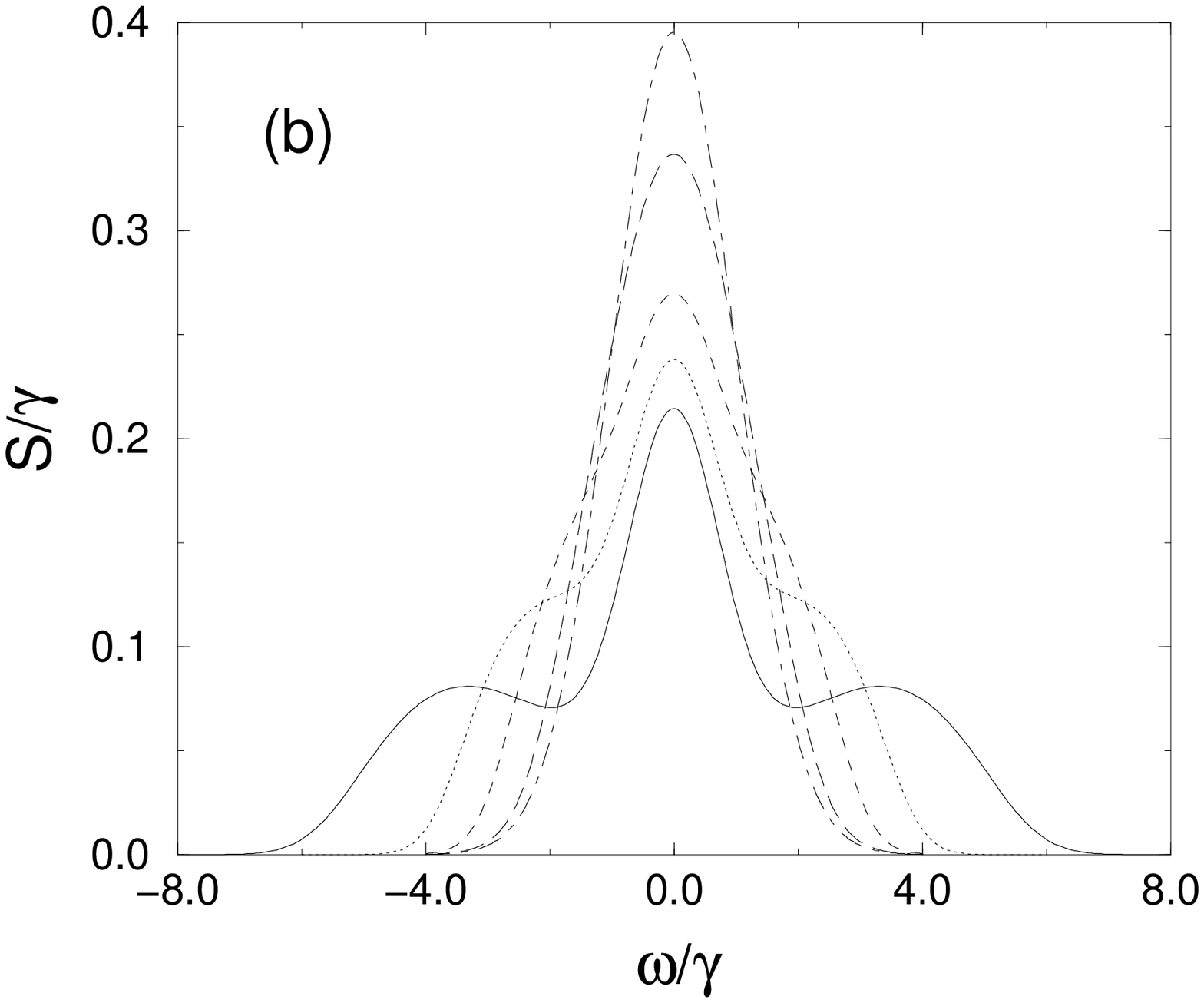}
\end{figure}

\vspace{70mm}

\centerline{FIG. 1}

\clearpage

\begin{figure}[htb]
\epsfxsize=6.5in
\epsfbox{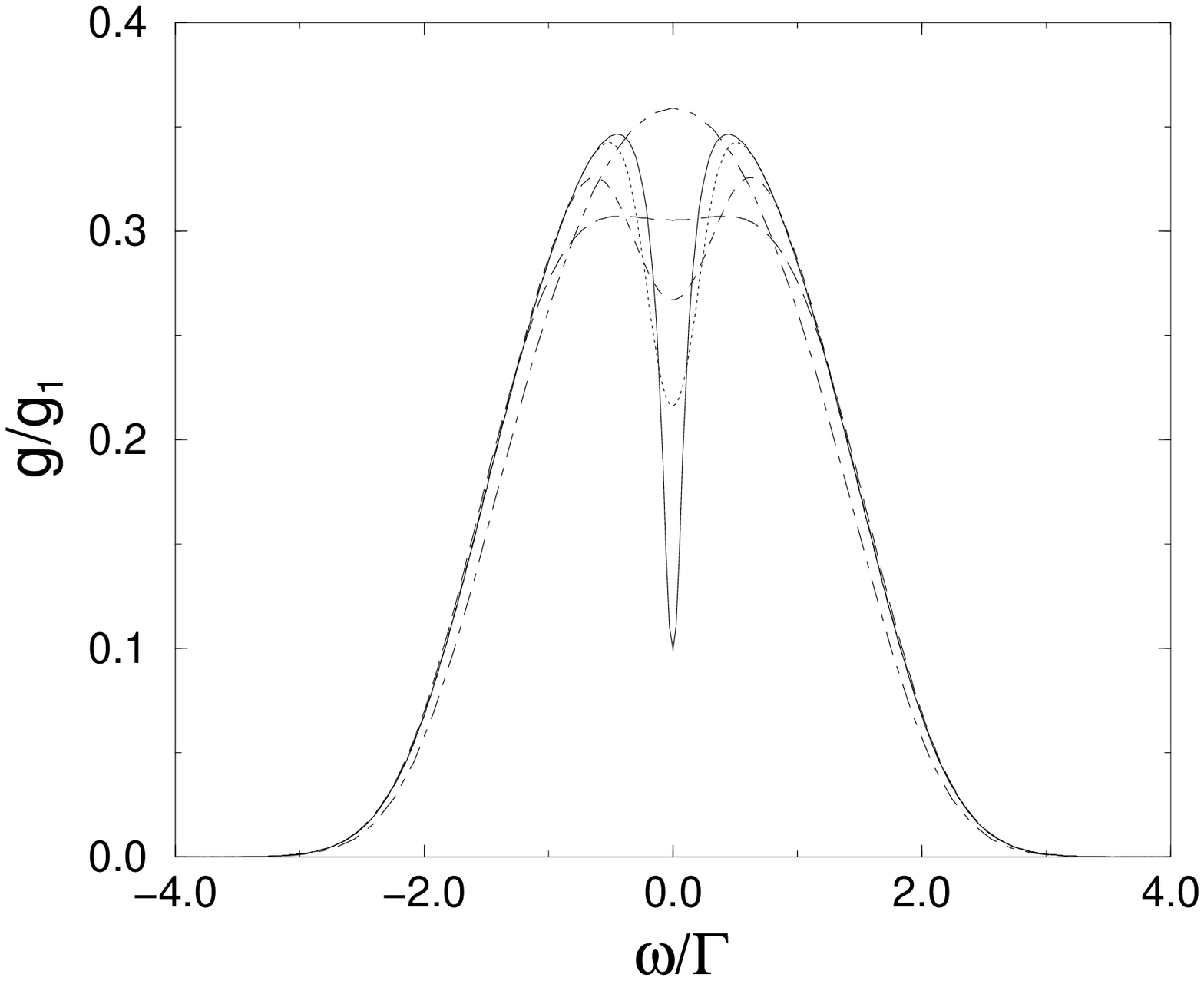}
\end{figure}

\vspace{70mm}

\centerline{FIG. 2}

\clearpage

\begin{figure}[htb]
\epsfxsize=6.5in
\epsfbox{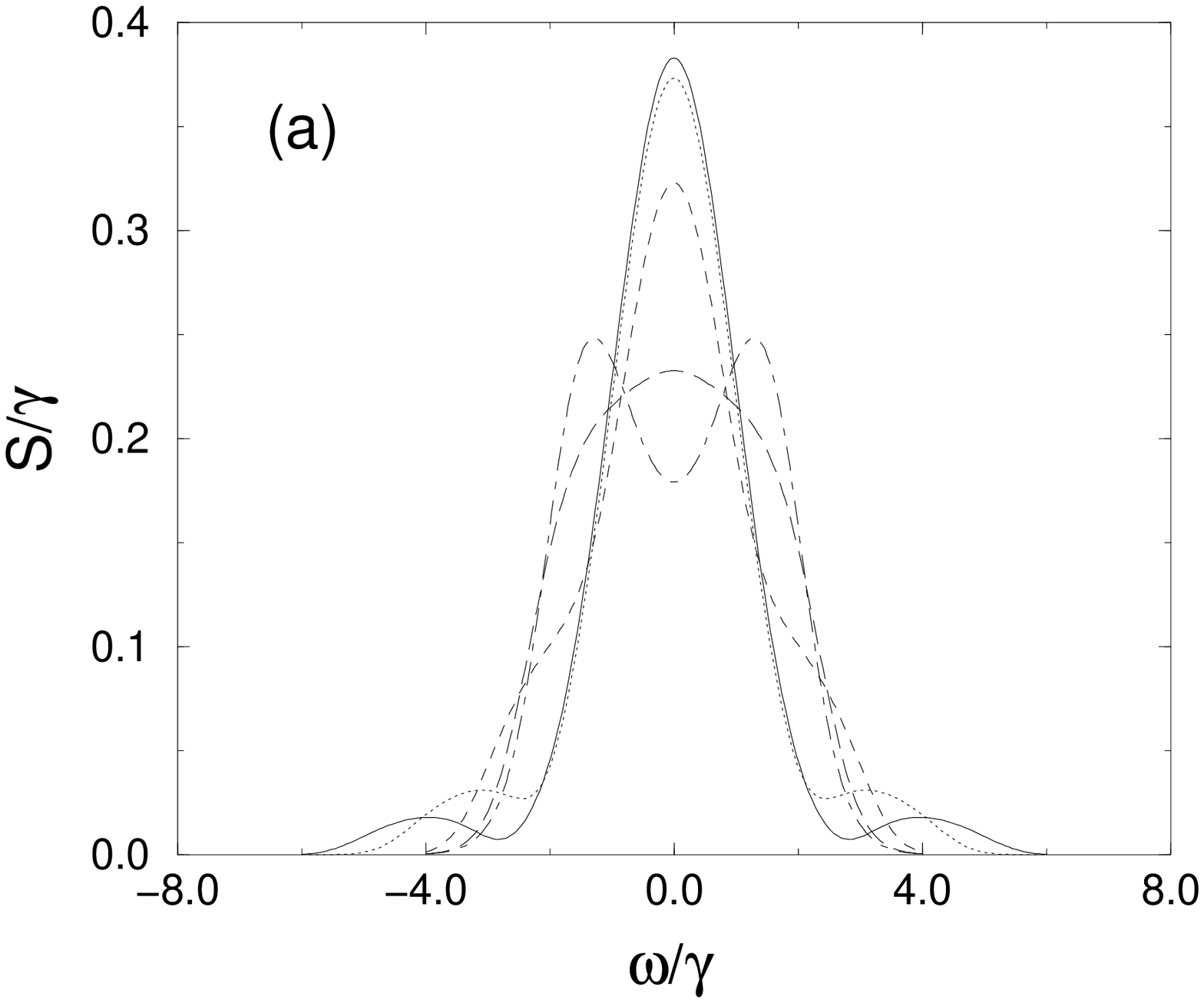}
\end{figure}

\vspace{70mm}

\centerline{FIG. 3}

\clearpage

\begin{figure}[htb]
\epsfxsize=6.5in
\epsfbox{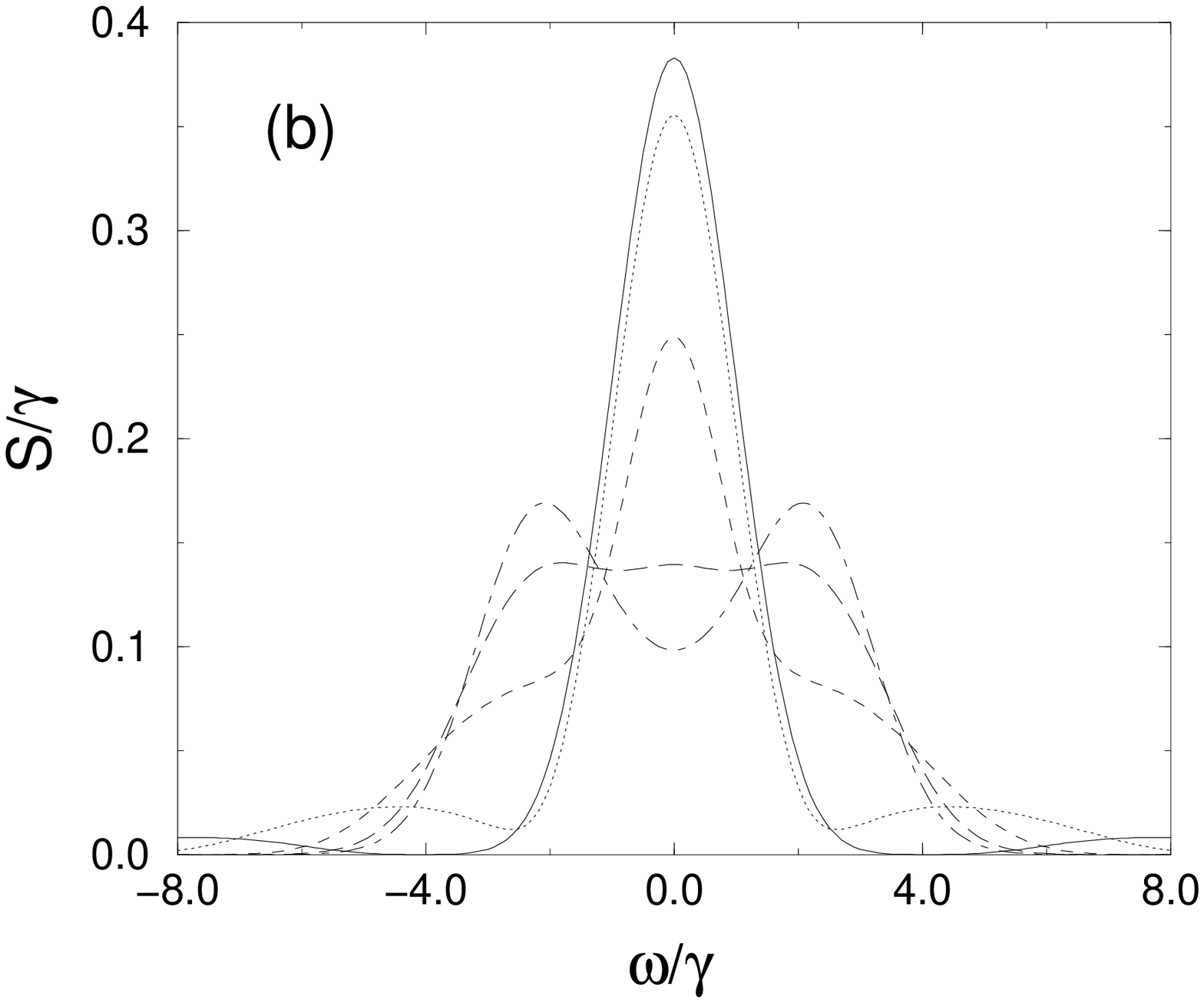}
\end{figure}

\vspace{70mm}

\centerline{FIG. 3}

\clearpage

\begin{figure}[htb]
\epsfxsize=6.5in
\epsfbox{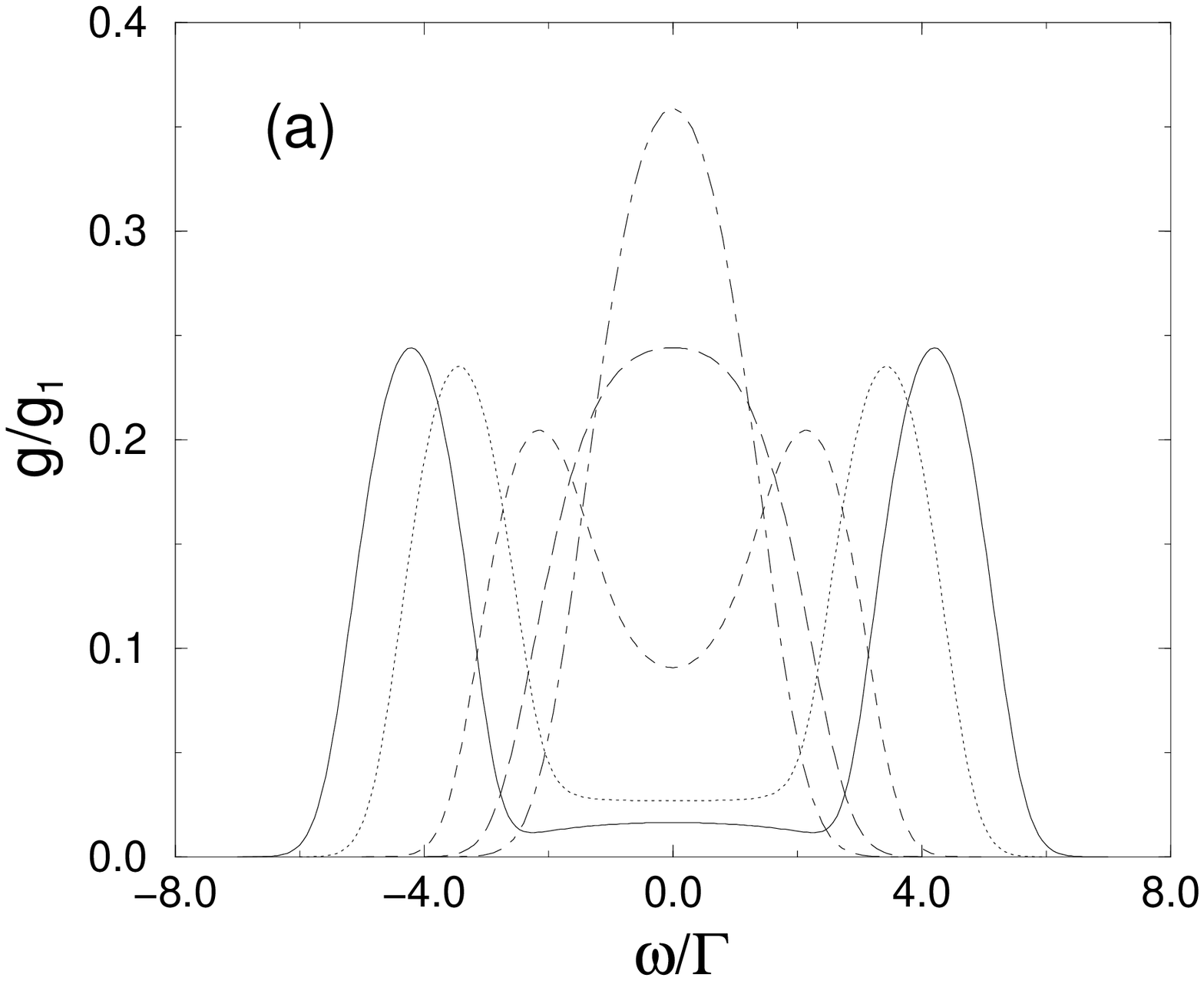}
\end{figure}

\vspace{70mm}

\centerline{FIG. 4}

\clearpage

\begin{figure}[htb]
\epsfxsize=6.5in
\epsfbox{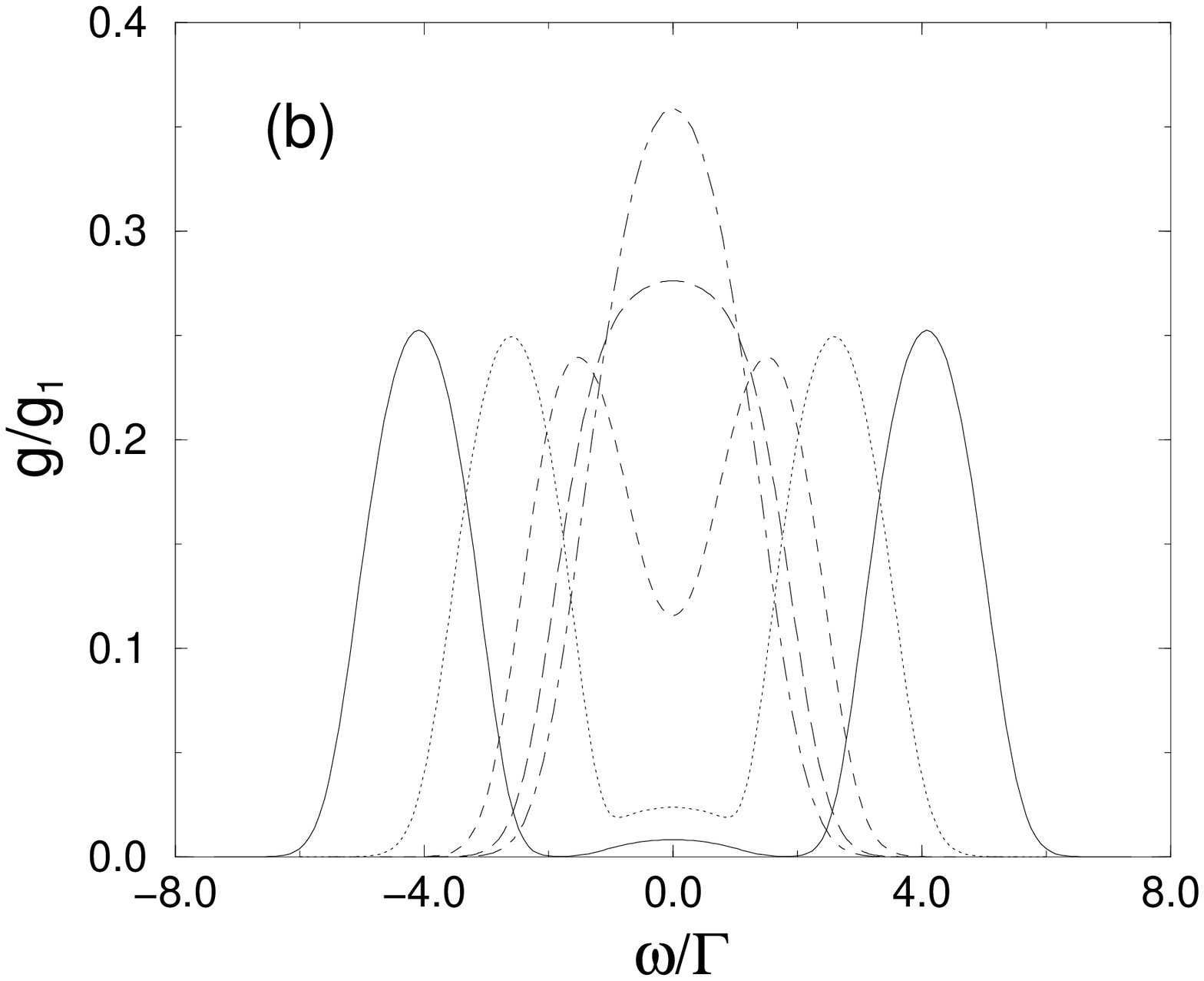}
\end{figure}

\vspace{70mm}

\centerline{FIG. 4}

\clearpage

\begin{figure}[htb]
\epsfxsize=6.5in
\epsfbox{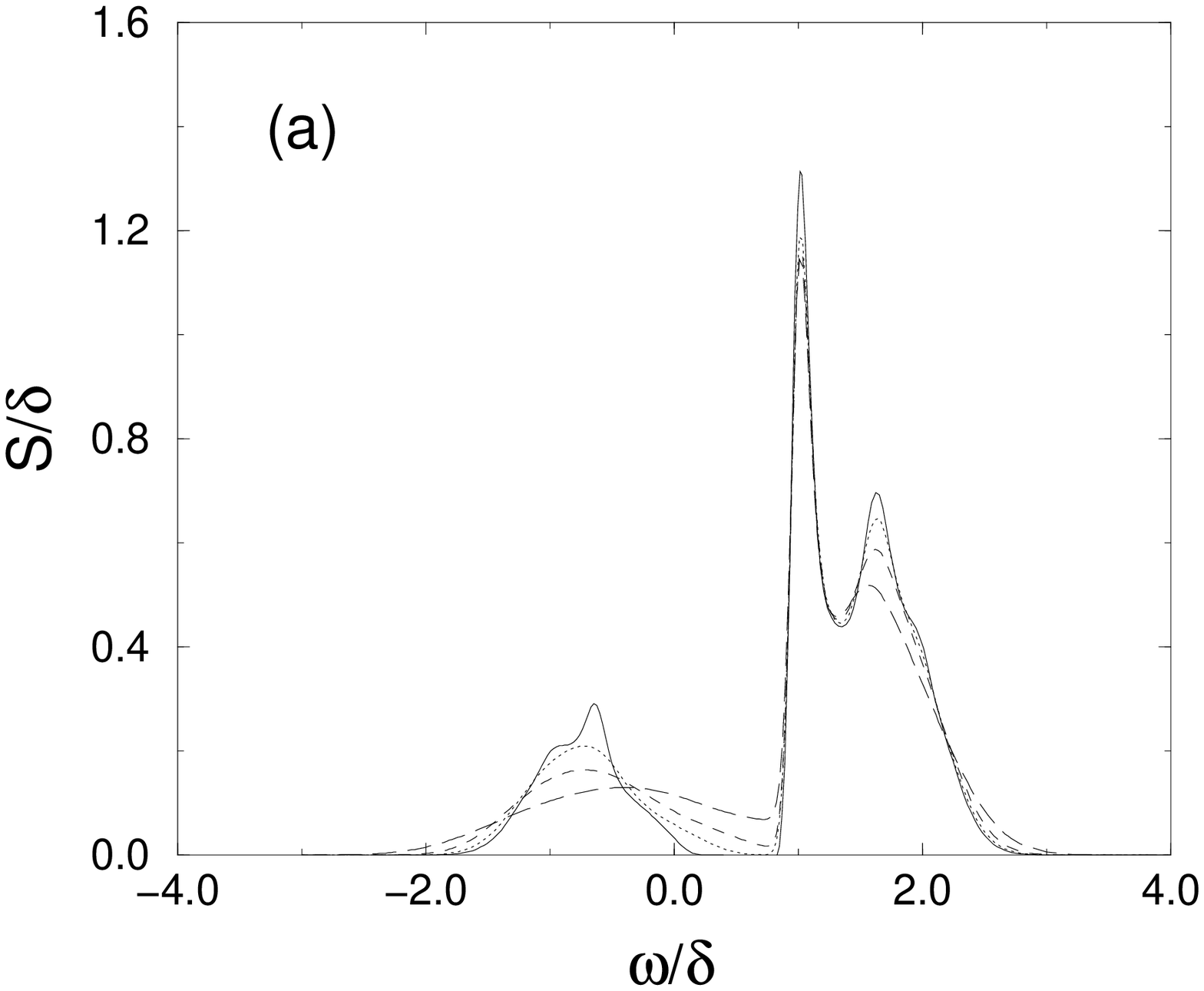}
\end{figure}

\vspace{70mm}

\centerline{FIG. 5}

\clearpage

\begin{figure}[htb]
\epsfxsize=6.5in
\epsfbox{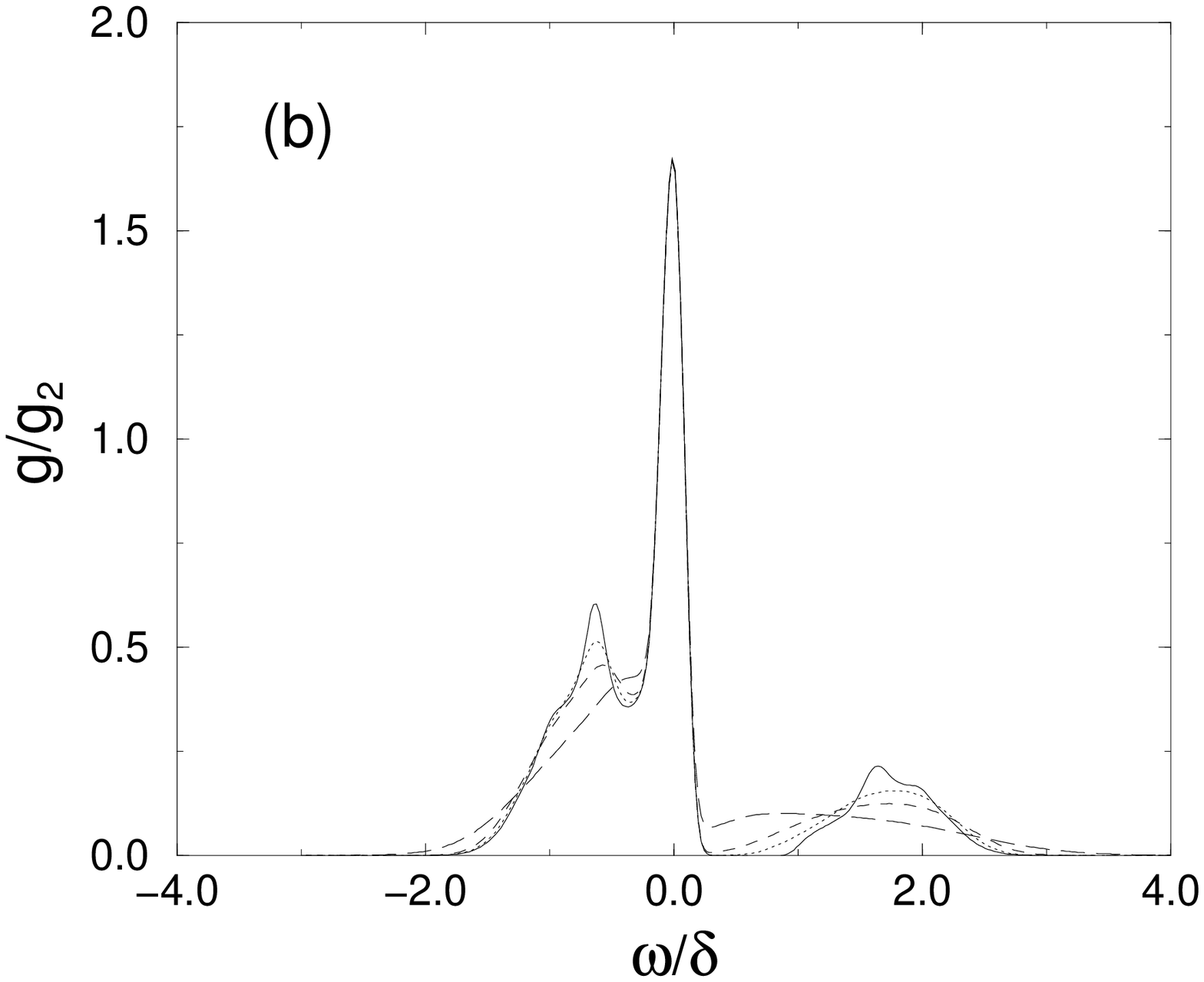}
\end{figure}

\vspace{70mm}

\centerline{FIG. 5}

\clearpage


\begin{references}
\bibitem{remark1}Eq.~(\ref{exp}) can be obtained with the help of the
identity   
\bea
1-e^{-i\beta\de^2/(\om-\ep)}=
\int_0^{\infty}dt e^{i(\om-\ep)t^2/4\beta\de^2}J_1(t),
\nonumber
\eea
where $J_1(t)$ is the Bessel function.
\bibitem{remark2}Here one uses the identity
\bea
\int_0^{\al} {d\beta\over\beta} \left(1-e^{-\beta}\right)=
C+\ln\al+\int_{\al}^{\infty}{d\beta\over\beta} e^{-\beta},
\nonumber
\eea 
where $C$ is the Euler constant.
\bibitem{remark3}The $\nu$--dependent constants in Eqs.~(\ref{Sgap})
and (\ref{ggap}) are the same as in the Appendix of Ref.~\onlinecite{G}
\end{references}
\end{document}